\documentclass[doublecol]{epl2}

\usepackage{amsmath,amssymb}
\usepackage[dvips]{epsfig}

\usepackage[latin1]{inputenc}
\usepackage[english]{babel}
\usepackage{ulem}
\usepackage{xcolor}

\newcommand{\red}[1]{{\color{black} #1}}

\title{The role of noise and advection in absorbing state
phase transitions}

\author{C. Barrett-Freeman, M. R. Evans, D. Marenduzzo, J. Tailleur}
\institute{SUPA, School of Physics and Astronomy, University of
Edinburgh, Mayfield Road, Edinburgh EH9 3JZ}


\abstract {We study the effect of advection and noise on the field
theory for directed percolation (DP). We show that even a very small
advective velocity is enough to change the universality class of the
dynamic phase transition. When the noise is taken to be proportional
to the square root of the population density, we find an additional
nonequilibrium ``spinodal'' line separating a region where an
exponentially decreasing density is metastable, from another one in
which it is unstable. If the noise is instead linear in the density,
the phase diagram changes dramatically both quantitatively and
qualitatively, and the spinodal line becomes a true phase
boundary.  We briefly discuss possible applications of our results to
microbial sedimentation and population dynamics in rivers.}

\begin{document}

\maketitle

The concept of universality classes has been employed with much
success in equilibrium statistical physics, gathering myriad phase
transitions into a handful of classes. The effort to extend and apply
this concept to nonequilibrium systems is still ongoing, and is one of
the foremost challenges of statistical physics.  Probably the widest
and best characterised nonequilibrium universality class to date is
that of directed percolation (DP)~\cite{hin}, a system often described
by the following Langevin equation\cite{jans}:
\begin{equation}\label{DP}
  \partial_t \rho=D \partial_{xx} \rho+a\rho-b\rho^2+\sqrt{\rho}\eta
\end{equation}
Here, $\rho=\rho(x,t)$ is a density, $D$ a diffusion coefficient, $a>0$ a
``growth rate'', $b>0$ a saturation constant, and $\eta(x,t)$ is
Gaussian white noise of unit variance.

DP is an archetypal nonequilibrium phase transition, which takes an
active phase, where the fluctuating density is non-zero,
continuously into an absorbing one with zero density. It succesfully
describes systems as diverse as chemical reaction-diffusion processes,
epidemic spreading, percolation through porous media, growing
microbial population and branching-annihilating random
walks~\cite{munoz}. This broadness is now expressed by the ``DP
conjecture'', stating that {\it all} systems exhibiting a continuous
transition into a unique absorbing state, characterised by a
one-component order parameter, and not showing any extra symmetries or
conservation laws, belong to the DP universality
class~\cite{jans,grass2}.  The DP conjecture has proven extremely
robust to changes in
the microscopic dynamics.  This brings us to the subject of this
article: what is the fate of the DP universality class in the presence
of advection?

One might expect that the addition of an advection term to (\ref{DP})
could be transformed away by a Galilean
transformation~\cite{saar,DNS00} thus rendering it irrelevant. This is
however not true in the presence of fixed boundaries.  For \red{most}
equilibrium systems boundaries are not expected to affect the critical
behaviour, however in a nonequilibrium system boundaries are known to
play a crucial role in determining phase
transitions~\cite{Mukamel}. To illustrate this and gain insight into
our question, we consider the effect of advection on the noiseless
limit of Eq.~\ref{DP}, which is nothing but the celebrated
Fisher-Kolmogoroff (F-KPP) equation~\cite{saar}. It clearly admits two
steady state solutions ($\rho=0$, $\rho=a/b$) and exhibits the
well-known Fisher Waves: wavefronts emerge from portions of the system
in the high density state $\rho=a/b$ and propagate into empty regions
with velocity $v_f=2\sqrt{Da}$~\cite{saar}.  It has recently been
shown~\cite{bs,DS07} that in the presence of boundaries, and upon
addition of an advection term $v \partial_x \rho$, F-KPP exhibits a
new low-density steady-state: an exponential phase in which the
density profile decays exponentially away from the boundary. The
competition between the advancing Fisher wave and the advection term
triggers a novel discontinuous non-equilibrium phase transition
separating this low-density phase ($v>v_f$) from the high-density one
($v<v_f$), in which $\rho\simeq a/b$ throughout the system. Restoring
noise and returning to DP, it is natural to question the robustness
of the exponential profile. Indeed, we now have two candidates for the
low-density phase: the exponential one or the usual absorbing state of
DP.

In this work, we show that in the presence of advection, the
low-density phase is always absorbed at long times whereas the fate of
the exponential state depends on the strength of the noise; it
is never seen at large noise but is metastable below a certain
threshold. The role played by the noise in this dynamical
transition is crucial.  We explore  this issue
further by considering noise proportional to the density rather than
to its square root and show that this transforms the dynamical
transition into a true phase transition.  Most importantly, we find
that the addition of an advective term to DP, however small, always
changes the nature of the transition, rendering it discontinuous. In
other words, advection is a relevant perturbation in renormalization
group jargon, in a similar way that, for instance, a magnetic field is
a relevant perturbation in an Ising model in equilibrium statistical
mechanics.

\begin{figure}
 \centering
 \includegraphics[width=6.cm]{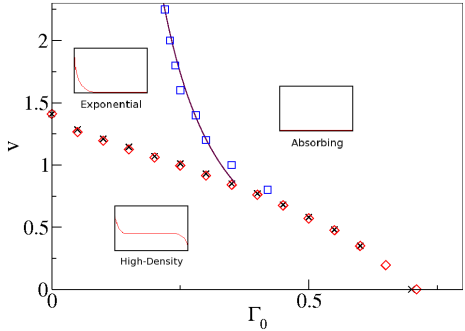}
 \caption{Phase boundary for $a=0.5$ and $\eta\propto\sqrt{\rho}$.
   The critical points found via the Dickman and Dornic \textit{ et
   al.}  algorithms are shown as black crosses and red diamonds,
   respectively. Both methods agree up to an accuracy of about
   1\%. \red{The solid maroon line correspond to the theoretical
   prediction (see below) for the spinodal line whereas the blue
   squares correspond to its numerical counterpart (see below).} }
 \label{pd1}
\end{figure}

\red{Our starting point is the following extension of Eq.~\eqref{DP}}
\begin{eqnarray}\label{full}
 {\partial_t \rho}  = 
  D{\partial_{xx} \rho}+
  v{\partial_x\rho}+a\rho  -  b\rho^2+\Gamma_0 g(\rho) \eta
\end{eqnarray}
\red{Our study is entirely carried out in one dimension; extension to
higher dimensions would be interesting but, since the advection term
can be extended in various ways, we leave this for further study.}  We
consider two different kinds of nonequilibrium noise distinguished by
the functional form of $g(\rho)$. In the first part of the article, we
choose $g(\rho)=\sqrt{\rho(x,t)}$, as in conventional DP (see
Eq.~\eqref{DP}), and refer to this as ``square root'' noise. Such
noise typically arises from fluctuations among individuals in a finite
population of average density $\rho(x,t)$. In the second case, that we
address toward the end of the paper, we consider a ``linear'' noise,
defined by $g(\rho)=\rho(x,t)$, which is usually met in the context of
population dynamics to model fluctuating
environments~\cite{may}. \red{This kind of noise is sometimes called
``multiplicative'' in the literature~\cite{munoz}, but we use this
terminology here to describe generic dependence of the noise on the
density}. Once $g(\rho)$ is chosen, $\Gamma_0$ is a parameter
used to scale the strength of the noise relative to the other
terms. Throughout the paper, we impose `no-flux' boundary conditions
at the top and bottom of the system\footnote{\red{{\it i.e}, we take
$D\partial_x \rho+v \rho=0$ at $x=0,L$}}, whose size is~$L$. \red{As
is usual for semi-infinite systems, we first take the large time limit
before sending the system size to infinity.}

To obtain our numerical results for the square root noise, we used two
different algorithms. The main technical difficulty is to ensure that,
upon time discretization of Eq.~\eqref{full}, the multiplicative noise
does not lead to unphysical negative densities when $\rho$ is small
(which is the case, for instance, when we are close to the critical
point of DP at $v=0$). The first method we use was originally proposed
by Dickman~\cite{dickman}, and entails a symmetrical truncation of the
noise coupled with discretization of the density such that $\Gamma_0
\sqrt{\rho}\,\eta$ never exceeds $-\rho(x,t)$. The second is an
improved integration scheme proposed by Dornic {\it et al.}
in~\cite{chate}.  Instead of drawing a random Gaussian number at each
$x$ to be later on multiplied by $\Gamma_0 \sqrt{\rho(x,t)}$, this
algorithm draws directly from the probability distribution function
which solves the Fokker-Planck equation associated with the linear
part of the local Langevin equation at each $x$ (see Ref.~\cite{chate}
for details).  After this, the deterministic non-linear part of the
equation is evolved via a finite difference scheme. In this case, the
density is never negative by construction and numerical problems are
avoided. We checked that both schemes yield very similar results. In
all the simulations presented in this article, we use $a=0.5$, $b=1$
and $D=1$. Typical integration parameters with Dickman's algorithm are
$dx=0.1$, $dt=0.001$, $L=1000$ whilst varying $v$ and $\Gamma_0$. For
the Dornic {\it et al.} algorithm, whose generalisation to an
advection term is quite straightforward, we could choose larger $dx$
up to 0.5 and $dt$ up to 0.05, considerably reducing simulation time.

First, we map out the phase diagram in the $\Gamma_0-v$ parameter
plane (see Fig.~\ref{pd1}). A full line of critical points
$v=v_c(\Gamma_0)$ now links the two limiting cases of DP ($v=0$) and
F-KPP with advection ($\Gamma_0=0$). For small $\Gamma_0$ and $v$, we
observe a high-density phase, where the total mass $M$ of the system
is extensive with system size, whereas the rest of the phase diagram
is composed of low-density regions (whether the exponential profile or
the absorbing state) where $M/L\to 0$ as $L\to\infty$. 
The absorbing state can be accessed via fluctuations and is thus
favoured by larger values of $\Gamma_0$. As a consequence, the
critical velocity $v_c(\Gamma_0)$, above which the high density phase
is not observed, is a decreasing function of noise strength.

Interestingly, long time simulations of the system show that its
dynamical behaviour in the low-density phase is not uniform. The
exponential phase is long-lived for small $\Gamma_0$ and we refer to
this regime as the `low noise' case.  While holding the velocity
fixed, increasing $\Gamma_0$ has the effect of decreasing the average
lifetime of the exponential state, until we can barely see it; this is
the beginning of the `strong noise' case. As we shall see, this
threshold of $\Gamma_0$ corresponds to a (nonequilibrium) spinodal line
which separates the two dynamical regimes (See Fig.~\ref{pd1}). We now
explain how the critical points were obtained and discuss the nature
of the phase transition, detailing in particular the differences
between the low and strong noise regimes.

In order to identify the critical points we define the order parameter
$m= M/L$, where $M$ is the total mass in the system.  We compute the
average of $m$ \red{in the quasi-stationary state\footnote{This is
the steady state of the probability distribution conditioned on
survival.}} over a number of different simulations, and then plot this
average as a function of $v$, for fixed $\Gamma_0$. For finite
systems, the crossover between low- and high-density phases sharpen
when the system size is increased, but the \red{order parameter}
curves intersect at a well-defined \red{non-zero} value of $v$ (see
Fig. 2 and 3), which we take as the critical velocity $v_c$. This
scenario is indicative of a discontinuous transition: in the
thermodynamic limit, as $v \nearrow v_c$ the density in the
\red{stead state} approaches a non-zero value, whereas {for} $v
\searrow v_c$ it is \red{strictly} zero. Thus, for all non-zero values
of $v$, there will be a discontinuous ``jump'' at the transition
point; the system no longer belongs to the DP universality class.

\begin{figure}
 \centering
 \includegraphics[width=6.cm]{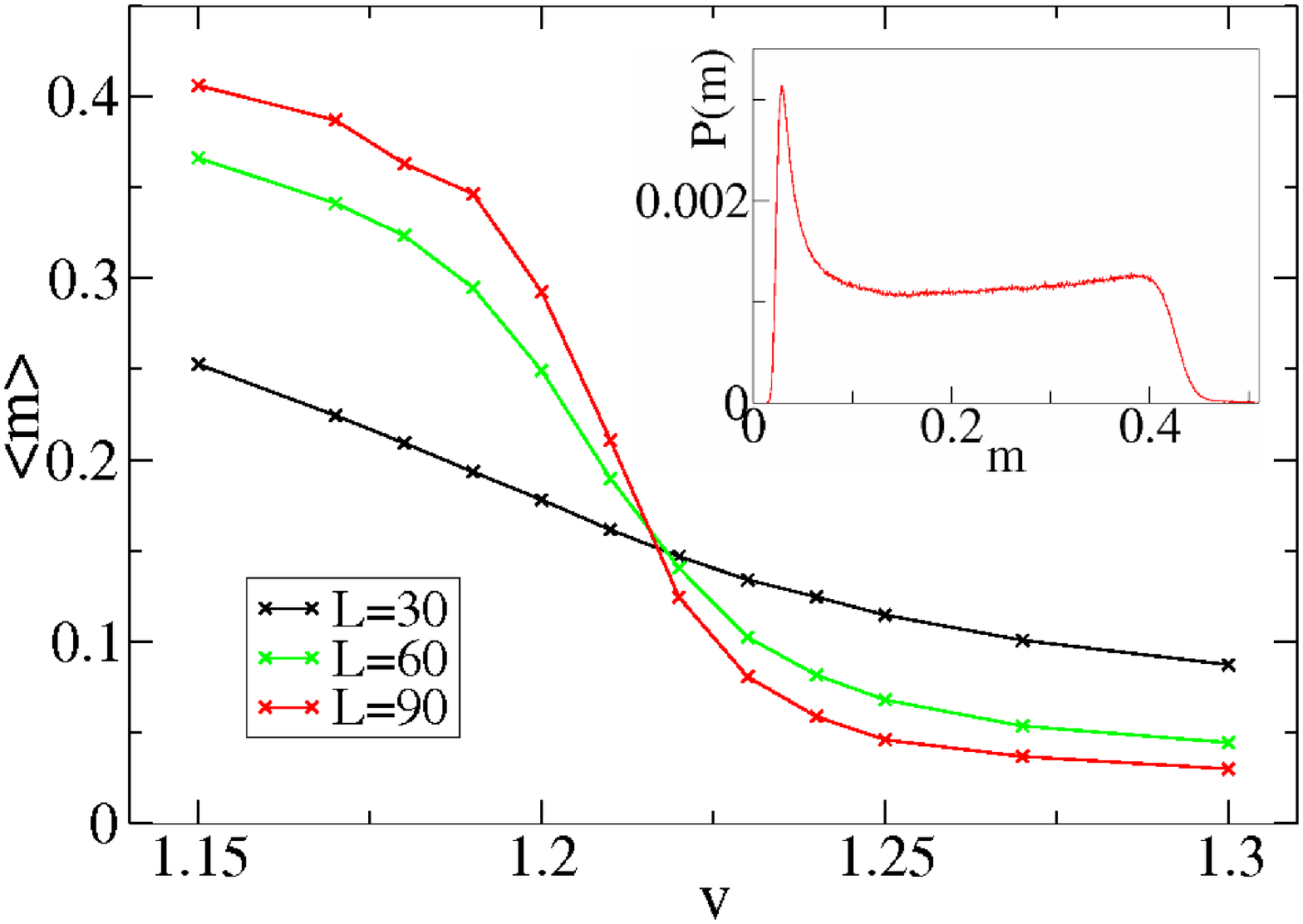}\\[1.cm]
 \caption{Plots of the average order parameter $\langle m \rangle$ as
   a function of $v$, for $\Gamma_0=0.10$ and for three different
   system sizes (see legend).  The presence of a crossing \red{of the
   order parameter curves} indicates that the transition is
   discontinuous. Inset: probability distribution of the order
   parameter, for $v=v_c\simeq 1.21$ and $L=100$. }
 \label{pdf}
\end{figure}

Whereas the nature of the transition differs from that of DP, it is
also different from the one of the noiseless case ($\Gamma_0=0$)
discussed in the introduction. In the absence of noise, a band
structure is formed at criticality and a stationary front separates
high- and low-density regions in the steady state. In the noisy case
considered here, the transition is in general different. Even the low
noise regime, while still bearing some resemblance to the noiseless
case, shows qualitatively distinct behaviour. This last point can be
appreciated by looking at the {quasistationary} probability
distribution of the average density of the system at criticality,
which is shown in the inset of Fig. 2. This distribution is composed
of an approximately flat part coexisting with a peak close to zero.
The physical interpretation is that in the low noise regime, the band
is still present but the front, no longer stationary, now performs a
random walk and occasionally gets stuck in the exponential profile. We
checked numerically that the roaming band indeed performs a random
walk by measuring the variance of the order parameter as a function of
time (data not shown).  The order parameter distribution is linked, in
equilibrium statistical mechanics, to the shape of the free energy.
In the case of a standard discontinuous transition, close to
criticality, there would be two coexisting free energy minima which
result in two peaks in the probability distribution of the order
parameter. In our case however, we have coexistence between one peak
corresponding to the exponential profile and a flat piece in which all
values of the total mass are essentially equiprobable. The latter
piece of the distribution corresponds to the high-density phase, in
which the average of the order parameter thus scales with the system
size.

\begin{figure}
 \centering
 \includegraphics[width=6.cm]{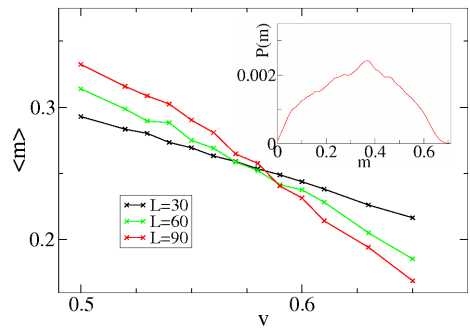}\\[1.cm]
 \caption{Plots of the average order parameter $\langle m \rangle$ as
   a function of $v$, for $\Gamma_0=0.50$ and for three different
   system sizes (see legend). The presence of a crossing \red{of the
   order parameter curves} indicates that the transition is
   discontinuous. Inset: probability distribution of the order
   parameter, for $v=v_c\simeq 0.58$ and $L=100$.}
 \label{large_noise}
\end{figure}

The transition into the low-density, large noise  regime is quite
different and, as might be expected, is reminiscent of what happens in
DP. In this regime, there is no longer any front, or band structure,
due to the large fluctuations, and the system now uniformly collapses
into the absorbing state. The numerics in this region are more
difficult. Since we are working with finite size systems, when
$v\lesssim v_c$ the system is occasionally  absorbed, even though it
properly belongs to the high-density phase. To overcome this
difficulty, we use PERM (Pruning and Enrichment Rosenbluth Method,
see~\cite{grass} for a recent review) for simulating the
quasistationary state. Despite the apparent similarity with DP, the
transition is still discontinuous, as can be seen in Fig. 3.
However, the order parameter distribution at criticality also differs
from the low noise case (see the inset of Fig. 3). First, the peak
corresponding to the exponential phase has disappeared. Then, the flat
part in the distribution is now replaced by a broad peak at a finite value.
For completeness, the quasistationary distribution, obtained using
PERM, should be complemented by a delta function at $\langle \rho
\rangle=0$. Therefore this case is closer to a standard discontinuous
transition, with two competing peaks at criticality. In the low
density phase, only the delta function survives in the
long-time limit.

As previously mentioned, the frontier between the low and strong noise
regimes can be probed via the dynamical stability of the exponential
phase. To do so, we study the evolution of the total mass in the
system $M(t)=\int_0^L\rho(x,t)dx$, obtained by integrating
eq.~(\ref{full}) with $g(\rho)=\sqrt{\rho}$:
\begin{equation}
  \label{eqnint}
  \dot M =a M-b \int_0^L \rho^2(x,t){\rm d}x +\Gamma_0\int_0^L
  \sqrt{\rho(x,t)} \eta(x,t){\rm d}x
\end{equation}
The diffusion and advection terms have dropped out due to the boundary
conditions. We approximate this by the following simplified Langevin
equation\footnote{A similar approximation was used by Munoz
in~\cite{munoz}}:
\begin{equation}
\dot M =a M-\beta M^2+\Gamma_0\sqrt{M}\tilde\eta(t)
\label{0D}
\end{equation}
where $\tilde \eta(t)$ is Gaussian white noise of unit variance. Both
noise terms in \eqref{eqnint} and \eqref{0D} are gaussian, have the
same mean and variance, and are therefore equivalent. The term $-\beta
M^2$ is an approximation---we know from \eqref{eqnint} that there must
be saturation terms in the effective dynamics of $M(t)$, and retain
only the lowest order in $M$. We believe this approximation to be
reasonable as we only consider the low-density regime, and hence small
mass. \red{(In the high-density phase, the mass would be extensive with
the system size and the approximation would break down.)} The parameter
$\beta$ contains the dependence on $v$ and $b$ but its form is not
known exactly.

\begin{figure}[ht]
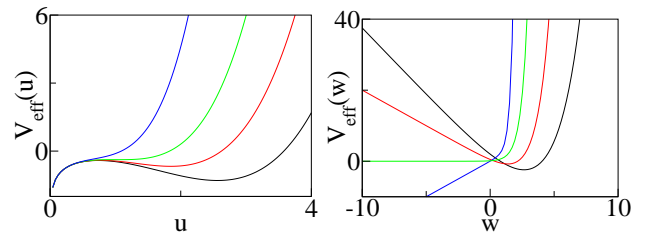

 \centering
 \includegraphics[width=4.cm]{fig4a.eps}
 \includegraphics[width=4.cm]{fig4b.eps}
 \caption{(a) Plot of the effective potential for
$\eta\propto\sqrt{\rho}$, $a=2$, $\beta=0.5$. From top to bottom, we
used $\Gamma_0=5.0;\, 2.8;\, 2.0;\, 1.5$ while $\Gamma_c=2\sqrt{2}$.  (b) Same
for $\eta\propto\rho$, $a=2$, $\beta=0.5$. From top to bottom, we used
$\Gamma_0=0.5;\, 1.0;\, 2.0;\, 4.0$ while $\Gamma_c=2$.}
 \label{v1}
\end{figure}

\red{The dynamical stability of the exponential is not easily studied
from \eqref{0D} as the noise is multiplicative. We therefore recast it
into an additive Langevin equation via the following change of
variable $u=2\sqrt{M}/\Gamma$. Using the Ito formula~\cite{Oksendal},
\eqref{0D} becomes
\begin{equation}
  \dot u = -\partial_u V_{\rm eff}(u) +  \eta;\quad V_{\rm eff}(u)=-\frac{a u^2}4 +\frac{\beta \Gamma^2 u^4}{32} +\frac {\log u}2
\end{equation}
The problem is now reduced to the diffusion of a particle in a
potential $V_{ \rm eff}$ at temperature $T=1/2$. A steady-state
solution is thus given by $P(u)\propto \exp[-2 V_{\rm eff}(u)]$. One
notes however that $P(u) \sim u^{-1}$ when $u\to 0$. The potential is
thus not normalizable, which simply stresses that the low density
phases are always absorbed as $t\to\infty$; the only normalizable
steady-state solution is $P(M)=\delta(M)$. The shape of $V_{\rm eff}$
nevertheless contains relevant information for the dynamics, as
illustrated in Fig. ~\ref{v1}a.} When
$\Gamma_0<\Gamma_c=\frac{a}{\sqrt{\beta}}$, the effective potential
has a local minimum corresponding to a potential well for positive
$M$. The well lies above the global minimum at $M=0$ and corresponds
to a metastable phase with finite mass: the exponential
phase. Conversely, for $\Gamma_0>\Gamma_c$ there is no metastable
state and the system falls directly into the absorbing state.
$\Gamma_c$ thus corresponds to a spinodal point at which the
exponential phase turns from metastable to unstable.

\begin{figure}[ht]
 \centering
 \includegraphics[width=6.cm]{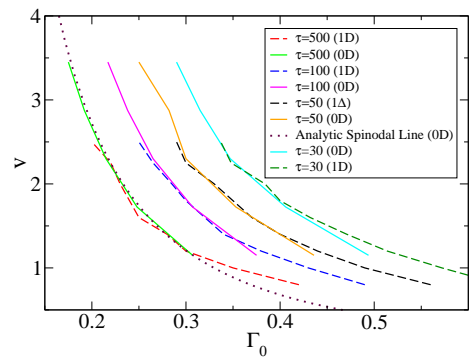}
 \caption{Contour lines in the $(v,\Gamma_0)$ plane for which $\tau$
equals, from left to right, 500, 100, 50 and 30. In order to plot the
contour lines of the 0D model, we use $\beta=2.3\, v$ \red{(see
below). The analytical prediction for the spinodal line is also
plotted, and closely matches the curves for which $\tau=500$.}}
 \label{spinodal}
\end{figure}

To construct the spinodal line $\Gamma_c(v)$ from the 0D model, we try
to relate $\beta$ to $v$ (the dependence on $b$ is not relevant since
$b$ is constant throughout our study). To do so, we compare the mean
time $\tau$ taken, for both the system and its 0D approximation, to
reach the absorbing state. We compute numerically\footnote{For the 0D
model, this mean first passage problem can be solved exactly up to a
numerical integration~\cite{politi}} a set of contour lines for $\tau$
in the planes $(v,\Gamma_0)$ and $(\beta,\Gamma_0)$, corresponding to
the 1D and 0D systems respectively (see Fig. 5).  Strikingly, simply
setting $\beta\simeq 2.3$ $v$ suffices to make the two sets of contour
lines overlap. This strongly supports the validity of the 0D model and
hence validates our interpretation of the frontier between low and
strong noise regimes as a dynamical phase transition. \red{We see on
Fig. 5 that the theoretical prediction for the spinodal line
corresponds to a mean first passage time (MFPT) to absorption of
$\tau=500$. In simulations, we indeed observe long-lived exponential
profiles when the MFPT to absorption is larger than $500$ whereas they
are barely seen otherwise.  The simulation data corresponding to
$\tau=500$ was thus used to pinpoint the spinodal on Fig. 1. }

\begin{figure}
 \centering
 \includegraphics[width=6.cm]{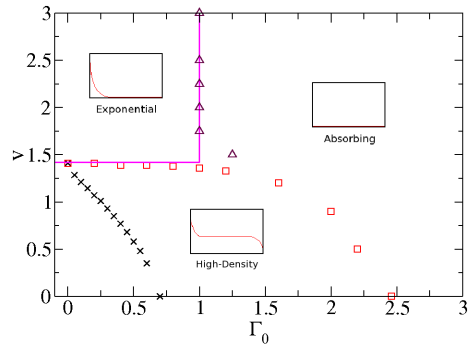}
 \caption{Phase diagram for $\eta\propto\rho$ ($a=0.5$). The red
   squares correspond to the transition between high- and low-density
   phases. Within the low-density region, there is now a second true
   phase transition between low and strong noise, indicated by the
   maroon triangles. For comparison, we include the phase boundary of
   the square root noise case (black crosses). The horizontal magenta
   line corresponds to the zero-noise transition point $v_c\simeq
   1.41$ whereas the vertical one is the theoretical prediction from
   the 0D model.}
 \label{pd2}
\end{figure}

We now turn to the investigation of the dependence of the phase
diagram on the type of noise used in eq.~\eqref{full}, by considering
$g(\rho)=\rho$.  Without advection ($v=0$), this model was studied
in~\cite{munoz,grinstein,tu} and it was shown that the system can be
mapped onto the KPZ equation, by means of a Cole-Hopf
transformation. As a consequence, the absorbing state phase transition
at $v=0$ is continuous, but does not belong to the DP universality
class. For instance, holding $\Gamma_0$ constant and varying $a$ close
to criticality yields $\langle \rho \rangle\sim (a-a_c)^{\beta}$ with
$\beta=1.5$, whereas $\beta\simeq 0.22$ in DP~\cite{dickman}.  In our
system, upon switching from the square root to the linear noise, the
phase diagram is dramatically altered (Fig.  \ref{pd2}).

First, the exponential profile is now completely stable for small
$\Gamma_0$ and the transition between low and strong noise is a true
phase transition. To understand why, we rely again on a 0D model,
obtained by replacing $\sqrt{M}$ by $M$ in the noise term of eq.
\eqref{0D}. \red{We this time consider the change of variable
$w=\Gamma^{-1} \log M$ to obtain an additive Langevin equation. Using
the Ito formula~\cite{Oksendal} the corresponding equation reads
\begin{equation}
  \dot w=-\partial_w V_{\rm eff}(w)+\eta;\quad V_{\rm eff}(w)=\Big(\frac \Gamma 2 - \frac a \Gamma\Big)w+\frac b {\Gamma^2} {\rm e}^{\Gamma w}
\end{equation}
and the putative steady-state is given by $P(w)\propto \exp[-2 V_{\rm
eff}(w)]$. For $\Gamma < \sqrt{2a}$, this is normalizable and the
system is thus not absorbed, there is a proper normalizable
steady-state distribution with $M\neq 0$. For $\Gamma \geq \sqrt{2a}$,
$\exp[-2 V_{\rm eff}(w)]$ is not normalizable and the steady-state
distribution once again corresponds to a delta function, the system
will be absorbed.} Considering now the stability of the exponential
profile, we see that the effective potential has a single minimum
which switches from \red{$w=-\infty$ ($M=0$) to finite $w$} depending
on the sign of the first term. \red{The transition point
$\Gamma_c=\sqrt{2a}$ corresponds to the normalization criterion, the
spinodal line has become a true phase transition.} Note that the
critical line $\Gamma_c(a,\beta)=\sqrt{2a}$ is now independent of
$\beta$ and hence of the velocity. Numerically, we find that, for
$a=1/2$, the transition line is almost vertical and very close to
\red{$\Gamma_c(v)=1$}. This close agreement with our predictions is
reinforced by the absence of any fitting parameter. As we approach the
high-density phase, the mass of the exponential profile increases, and
the slight bend of the transition line is presumably due to non-linear
terms beyond the quadratic one, neglected in the 0D approximation. In
the low noise regime, for $\Gamma_0<\Gamma_c$, the exponential state
is now stable rather than metastable and hence constitutes a true
phase.

\begin{figure}
 \centering
 \includegraphics[width=6.cm]{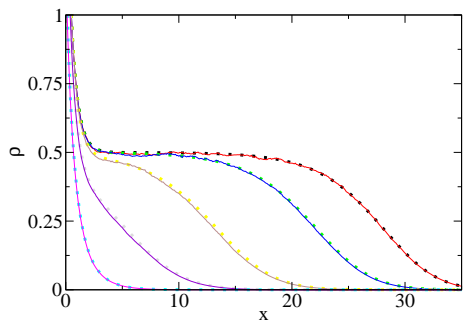}
 \caption{Plots of the steady state density profiles for the linear
noise ($\Gamma_0=0.1$, solid lines) and noiseless ($\Gamma_0=0$,
dotted lines) cases, for $L=40$ and values of $v$, from right to left,
$1.30$, $1.36$, $1.39$, $1.40$ and $1.41$.}
 \label{sn}
\end{figure}

Beyond the transformation of the dynamical transition into a true
phase transiton, the transition between the high- and low-density
phases (whether exponential or absorbing) is also changed
significantly. To pinpoint this critical line, we proceed numerically
as before. The model is now easier to simulate since close to the
transition, when $\rho\ll a/b$, the fluctuations remain of order
$\rho$ and are much smaller than in the previous case, where they
scaled as $\sqrt{\rho}$. Rare events leading to absorption are
unlikely and Dickman's algorithm works very well, in particular
discretization of the density is unnecessary. Our results suggest that
for small noise, fluctuations are irrelevant and one recovers a
transition identical to one of the deterministic limit---see
Fig.~\ref{sn}. Indeed, numerically the transition line between
exponential and high-density regimes is independent of
$\Gamma_0$---hence horizontal---with $v_c \simeq 1.41$ as in the
noiseless limit.  Note that this transition is discontinuous and the
band is stationary -- in sharp contrast with the $\sqrt{\rho}$ noise
where it performed a random walk at criticality. In the large noise
case, where the stable state is absorbing rather than exponential, our
simulations are consistent with a continuous transition (see Fig. 8)
and more data would be needed to discriminate it from the $v=0$
limiting case.

In conclusion, we have studied the effect of advection and the role of
noise in the context of continuous absorbing phase transitions, and
directed percolation in particular. Our main result is that as soon as
an advective velocity is added, the transition between the low- and
high-density phases becomes discontinuous and therefore is no longer
in the DP universality class. It is important to note that, in line
with other terms in DP, the advection term is a nonequilibrium term,
as it is related to the transport rather than to the thermodynamic
properties of the system. Furthermore, we have \red{illustrated} that
the type of noise used to model non-equilibrium phase transitions can
control the kind of transitions observed. For instance, switching from
square root to linear noise in our system transformed a dynamical
phase transition into a true phase transition, stabilizing a
previously metastable state. In addition to this, it appears that a
large portion of the critical line has changed from discontinuous to
continuous, although more extensive simulations would be required to
confirm this. This important quantitative and qualitative difference
\red{highlights once again~\cite{munoz}} that care should be exercised
when deriving fluctuating hydrodynamic equations for non-equilibrium
models, as the very form of the noise, which is sometimes overlooked,
may drive unexpected changes in the physics of the system.

\begin{figure}
 \centering
 \includegraphics[width=6.cm]{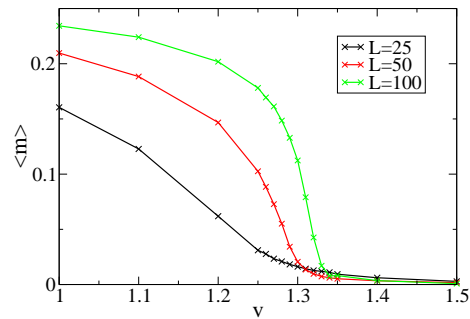}
 \caption{Plots of the average order parameter $\langle m \rangle$ as
   a function of $v$, for $\Gamma_0=1.25$ in the linear noise case,
   for three different system sizes (see legend). This figure is
   consistent with a continuous transition.}
 \label{ln2}
\end{figure}

We close with a brief mention of a possible application of our results
to experiments with microbial systems.  In population dynamics,
Eq.~\ref{full} at zero or small noise has been used to model bacterial
sedimentation in pipettes~\cite{bs}, plankton sinking in
oceans~\cite{huisman} (although in that work the logistic growth law
was substituted by a non-local forcing term), and the persistence of
populations of aquatic organisms in rivers~\cite{speirs} (i.e. the
problem of how organisms can resist being swept upstream by advection
from the flows in the rivers they inhabit). In all of these cases,
experiments suggest that the Fisher equation is a good, although
approximate, starting point to study these systems in the absence of
advection: therefore the regime of interest in our calculations is the
small noise one where a front can still be identified. Our results
suggest that in this situation, being at or (for finite size systems)
close to criticality brings with it a giant increase in fluctuations,
and as a result we would predict that the system under these
conditions would be unable to reach a steady state: the Fisher
wavefront would indefinitely perform a random walk instead. It would
be interesting to see whether controlled experiments in bacterial
colonies subject to, e.g., uniform flow in microfluidic devices may be
designed which give a transition between a persistent phase in which
the Fisher wave moves faster than the flow to another one in which the
bacteria are swept away by the moving fluid. If this is the case, our
results suggest that it would be extremely interesting to closely
monitor the behaviour of these experiments close to this transition.

This work has made use of the resources provided by the Edinburgh Compute and 
Data Facility (ECDF). ( http://www.ecdf.ed.ac.uk/). The ECDF is partially 
supported by the eDIKT initiative ( http://www.edikt.org.uk).
CBF is funded by EPSRC and JT acknowledges funding from EPSRC grant
EP/030173.

\end{document}